\documentclass{emulateapj}

\slugcomment{Accepted in ApJ, November 2006.}

\shorttitle{MOLECULAR SURVEY OF CRL 618}
\shortauthors{Pardo et al.}

\begin{document}


\title{Molecular Line Survey of CRL 618 from 80 to 276 GHz and complete model}


\author{Juan R. Pardo, Jos\'e Cernicharo \& Javier R. Goicoechea}
\affil{IEM - Departamento de Astrof\'{\i}sica Molecular 
              e Infrarroja, CSIC, Serrano 121, E-28006 Madrid, Spain. }
\email{cerni,pardo@damir.iem.csic.es}
\and
\author{Michel Gu\'elin} 
\affil{Institute de Radioastronomie Millim\'etrique, 30 rue de la Piscine, F-38406 Saint Martin d'H\`eres, France}
\and
\author{Andr\'es Asensio} 
\affil{Instituto de Astrof\'{\i}sica de Canarias, E-38205, La Laguna,
       Tenerife, Spain}


\begin{abstract}
We present the complete data set, model and line identification 
of a survey of the emission from the 
C-rich protoplanetary nebula CRL 618 performed with the IRAM-30m telescope in 
the following frequency ranges: 80.25-115.75 GHz, 131.25-179.25 GHz, and 
204.25-275.250 GHz. A selection of lines from different species has been 
used in previous works to derive the structure of the source, its physical 
conditions and the chemical abundances in the different gas regions. In this 
work, we have used this information to run a global simulation of the spectrum 
in order to check the consistency of the model and to ease the task of 
line identification. The total number of lines that have a correspondence in 
both data and model is $\sim$3100, although quite often in this object many 
lines blend into complex features so that the model, that takes into account 
line blending, is a key tool at this stage of the analysis. Of all the lines 
that we have been able to label, $\sim$55\% of them belong to the different 
forms of HC$_3$N, and $\sim$18\% to those of HC$_5$N. The density of remaining 
unidentified features above the 3$\sigma$ limit is only one per $\sim$2.1 GHz 
(74 features), which is unprecedented in the analysis of this type of large 
millimeter-wave line surveys.
\end{abstract}



\section{Introduction}

In order to study the evolutionary stages from Asymptotic Giant Branch (AGB) 
to Planetary Nebulae (PN), multiwavelength and multi-object data are necessary to reveal 
the key changes in physical conditions and chemical composition as completely as 
possible. Even so, the task is not trivial due to the many distinct environments that can 
be present in these C-rich or O-rich objects including: very fast winds and presence of 
shocks (Cernicharo et al. 1989; 
Davis et al. 2005; Bujarrabal et al., 2002),  
binarity (S\'anchez-Contreras et al. 2004, De Marco et al. 2004), magnetic fields 
(Frank \& Blackman 2004; 
Huggins \& Manley 2005), late thermal pulses (Asplund et al. 1999, Pavlenko et al. 2004), 
etc... Molecular 
spectroscopy at millimeter, submillimeter and FIR wavelengths is a useful tool to study the early stages of this 
evolution because large carbon chain and metal containing molecules form in the atmospheres of late-type stars. 
This rich formation chemistry 
exists in C-rich objects due to the relatively high densities and the presence 
of dust and shocks created by significant mass loss (Cernicharo 2004). In addition, 
the potentially photodissociating radiation from the 
central star, when it changes its spectral type, is blocked in a compact inner region. As 
the evolution goes on, the shielding of high energy photons from the 
central star, that has evolved toward types A, B and O is less effective so that the 
molecular content in the envelope decreases greatly. 

The ideal targets to study the formation chemistry of large carbon-chain molecules  
are late type stars with significant mass loss (e.g., IRC+10216), protoplanetary nebulae 
(e.g., CRL 2688 and CRL 618) and 
very young planetary nebulae (e.g., NGC 7027). Herpin et al. (2001), performed a study using 
several CO lines across the mm and submm ranges in the latter four objects, taken as prototypes 
for the different steps of the AGB to PN evolution, aimed at determining the physical 
conditions prevailing for the molecular gas. S\'anchez-Contreras et al. (2002, 2004) have 
undertaken a multiwavelength study of CRL 618. Cernicharo et al. (1996, 2001a,b) have 
explored the ISO data of CRL 618 and 
IRC+10216 with interesting findings such as the first evidence for the first aromatic molecule seen 
outside the solar system (C$_{6}$H$_{6}$ in CRL 618). Remijan et al. (2005) have performed 
a search for large molecules toward CRL 618 although no detection of biologically 
important molecules could be reported. Finally, Cernicharo 
et al. (2000) published a $\lambda$=2 mm survey of IRC+10216. Therefore, our motivation 
for the complete {\it Institute de Radioastronomie Millim\'etrique} (IRAM) 30 meter telescope 
survey of CRL 618, fully presented here, has been to gather the 
most complete molecular information on one particular stage (protoplanetary nebula, PPNe) 
of low-mass stellar evolution, of which CRL 618 is the best C-rich 
example. The data on their own have allowed a quite precise determination of physical 
conditions and molecular abundances in this object (Pardo et al. 2004, 2005; 
Pardo \& Cernicharo 2006; hereafter P04, P05 and PC06). These previous works have allowed 
us to build a complete model to be compared with the whole data set, thus simplifying the 
task of isolating U-lines during the line identification process.
 
\section{Observations Summary}
\label{sect:obs}

The CRL 618 IRAM-30m line survey was completed in 1994-2002. In order to keep the 
calibration of the spectra as consistent as possible, the observing procedure 
was kept unchanged through the years. Nevertheless, technical works at the telescope 
and varying atmospheric conditions are responsible for a relative calibration uncertainty 
of the order of 10\%. This uncertainty in the calibration was based on observing the same 
lines at different epochs. The analysis of more than 15 lines in different vibrational 
states of HC$_{3}$N has allowed to notice a few scans with obvious calibration 
problems (less than 10 over more than 350 spectra, or $<$3\%) most probably due to an 
erroneous calibration scan. Those few spectra were re-scaled based on the overall results 
of the rotational HC$_{3}$N ladders.  

In order to avoid signal and image sideband confusion, 
image sideband rejection larger than 12 dB was used. Features that could be 
attributed to sideband contamination are only a few, generally related to 
a very intense line (CO, HCN, HNC, HCO$^{+}$,...) in the image side band. The pointing 
and focus were always checked on the target source and the peak level of the 
gaussians after the last correction of each pointing session has been used to 
study the spectral behavior of the millimeterwave continuum in CRL 618. The 
pointing itself was kept within 2'' accuracy. The observations were carried 
out using the wobbler-switching mode with offsets of 60'' and 
frequencies of 1 Hz, in order to obtain very flat baselines.  
The backends were two 512$\times$1 MHz filter banks connected to the receivers 
operating below 200 GHz, and an autocorrelator with channel widths of 
1.25 MHz connected to the receivers operating in the 1.3 mm window. System 
temperatures were typically 100-400 K at 3 mm, 200-600 K at 2 mm,  
and 300-800 K at 1.3 mm.

Complementary observations above 280 GHz have been carried out with the 10.4 m dish 
of the Caltech Submillimeter Observatory (Mauna Kea, Hawaii) and some of them have 
been used in the analysis presented in P04, P05 and PC06. The complete results 
above 280 GHz will be published elsewhere.

\section{Results}

In order to illustrate the overall results 
of the survey we include here two tables (\ref{table-species} and \ref{table-species-bis}) 
describing the molecules detected and the $J_{up}$ 
or energy range in which the lines of a particular species are seen. Other species included 
in the model, although marginally or not detected, appear also in the same table. We also provide 
four figures showing the overall data and model in four spectral ranges, corresponding to those 
of the available receivers at the IRAM-30m telescope: 3 mm, 2 mm, 1.3 mm (lower end) and 
1.3 mm (upper end). In order to have an idea in the printed version of the correspondence between 
data and model, we present nesting zooms in frequency in all figures, so that the most detailed 
panel shows 0.5 GHz wide spectra (corresponding to the frequency coverage in each observational 
setting). The whole survey at this level of detail needs about 300 pages of figures and tables and 
therefore this is available only as on-line material. Finally, Table \ref{u-lines} provides the position 
and intensity of the remaining unidentified features after the analysis performed. 

\subsection{Continuum emission}
\label{sectcont}
 The continuum in  T$_{A}^{*}$ scale was derived from the pointing scans, 
as explained in section \ref{sect:obs}, after discarding a few obvious bad scans.  
T$_{MB}$ and fluxes have been derived from a fit of those data to a horizontal line, 
scaled by the corresponding beam and forward efficiency factors of the 30m telescope 
(B$_{eff}$ ranges from 0.79 to 0.42 and F$_{eff}$ from 0.95 to 0.88 in the frequency 
range 80 to 275 GHz for this telescope). This 
analysis is not too much affected by the presence of lines because the total 
flux from the lines is less than 3-5 \% relative to the continuum flux in most frequency 
settings. As a comparison, the total flux is dominated in fact by the lines 
in millimeter wave observations toward high mass star forming regions such as Orion 
($>$ 50\% of the total flux towards the IRc2 position at these wavelengths, Tercero et al. {\it in prep}). 
A few frequency settings with very strong lines such as the lowest rotational 
transitions of CO and HCN have been ignored for the continuum flux analysis in CRL 618. The results 
were presented and discussed in sections 4.1 and 5.1, Fig1, and Table 2 of P04. In terms 
of T$_{MB}$, the continuum follows a straight line from 0.52 K at 80 GHz to 1.52 K at 275 GHz. 
The spectral behavior seems to be basically the same in the frequency range 280-360 GHz according to  
our results obtained at the Caltech Submillimeter Observatory (P04, P05).

\subsection{Overall line spectra} 

\begin{figure*}[h]
 \centering
 \includegraphics[angle=0,width=.8\textwidth]{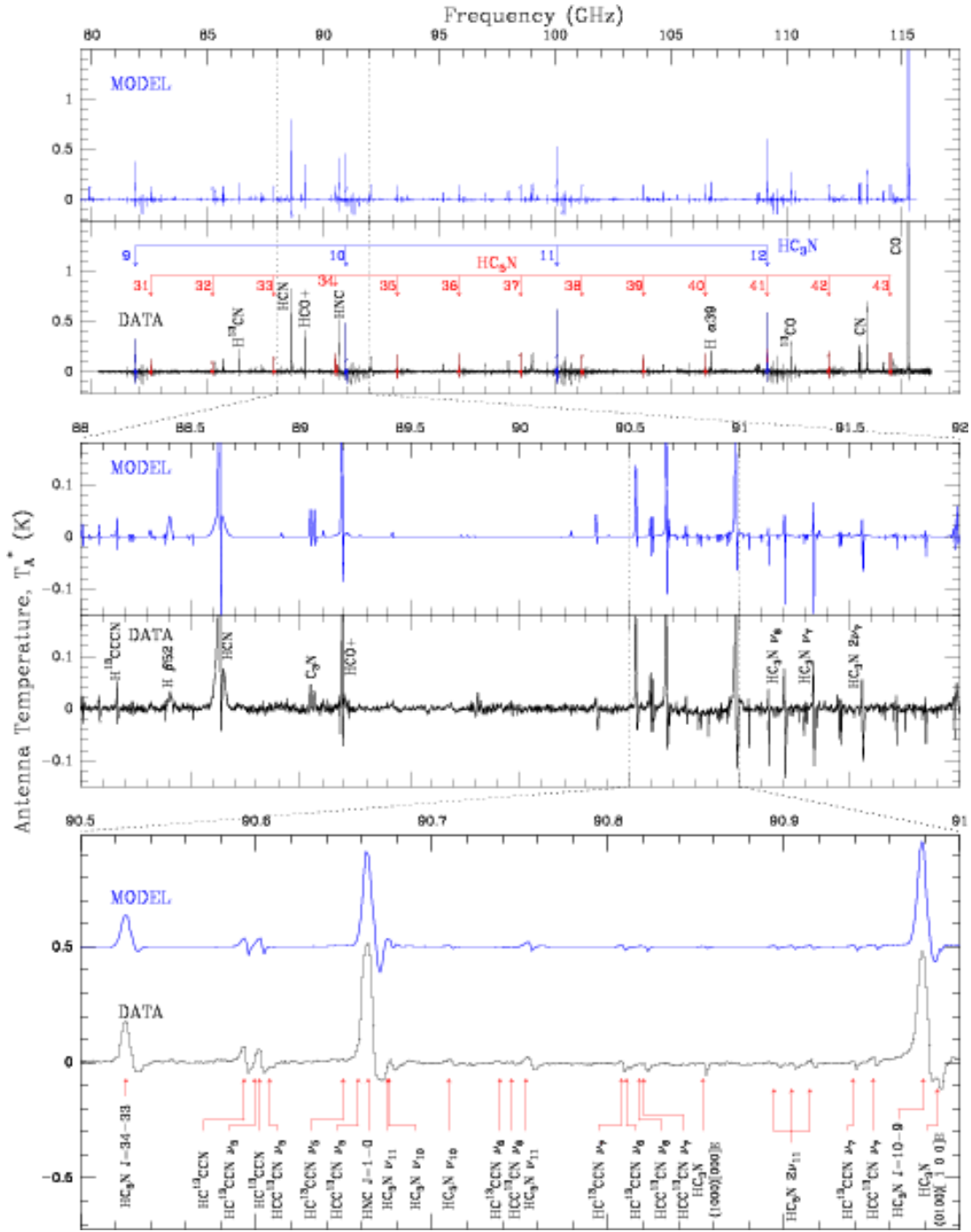}
 \caption{Results of the CRL 618 IRAM-30m survey in the frequency 
range 80.0-116.5 GHz (3 mm window) compared to the model devepoloped according 
to the results found in the series 
of papers P04, P05 and PC06. We include nesting 
zooms on intervals covering 4 GHz and 0.5 GHz respectively. The data are shown in terms of 
T$_A^*$ with the continuum level substracted. The labels next to the arrows 
in the upper panel indicate the J$_{up}$ number of the corresponding HC$_3$N or HC$_5$N 
transition. The complete set of detaliled 
figures at 0.5 GHz intervals with line identifications (similar to the lower panel), is available 
as on-line material only. }
              \label{Fig0a}%
    \end{figure*}

\begin{figure*}[h]
 \centering
 \includegraphics[angle=0,width=.8\textwidth]{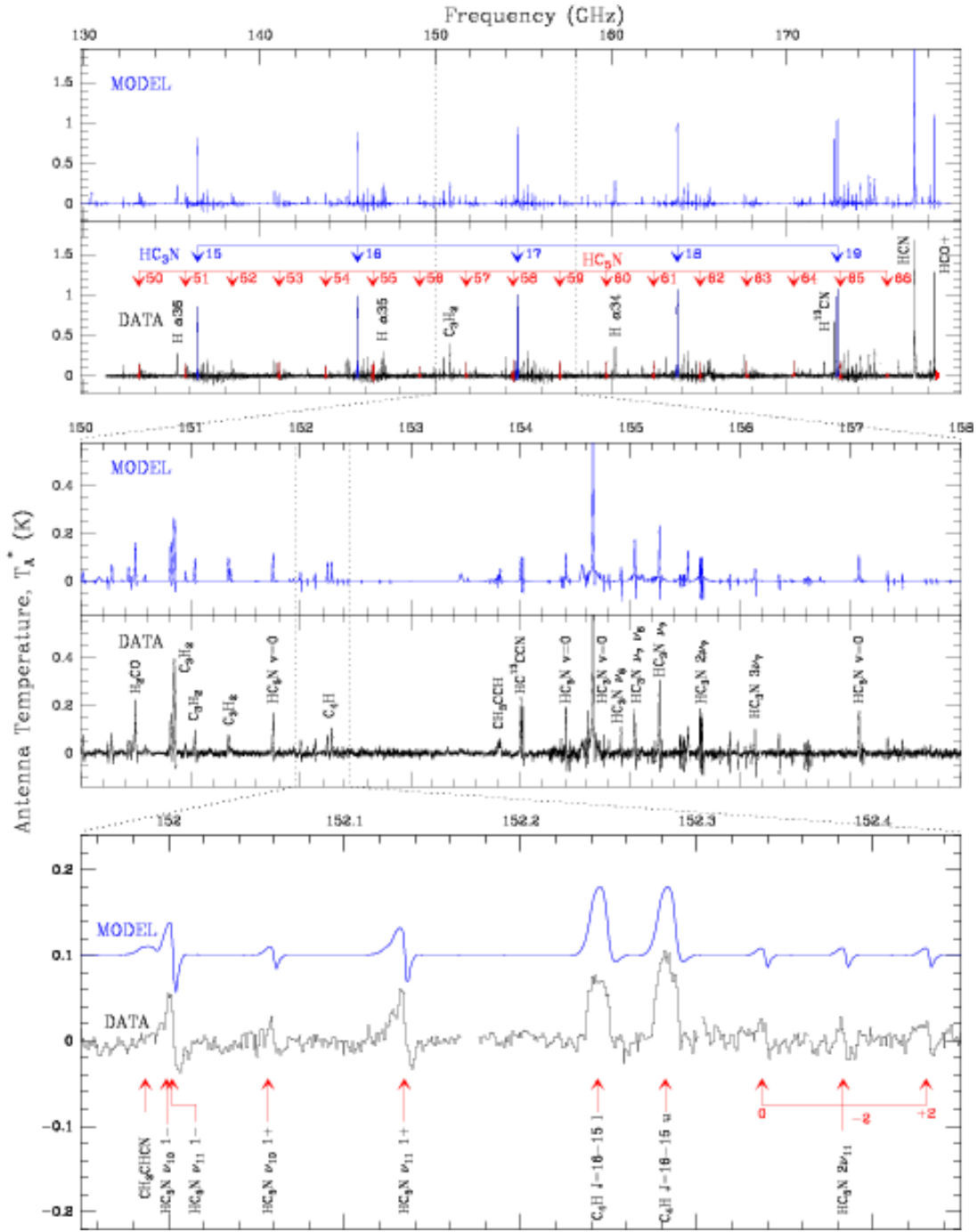}
 \caption{Same as previous figure but for the frequency 
range 130.5-179.0 GHz (2 mm window). The first zoom shows 8 GHz instead of 4.}
              \label{Fig0b}%
    \end{figure*}

\begin{figure*}[h]
 \centering
 \includegraphics[angle=0,width=.8\textwidth]{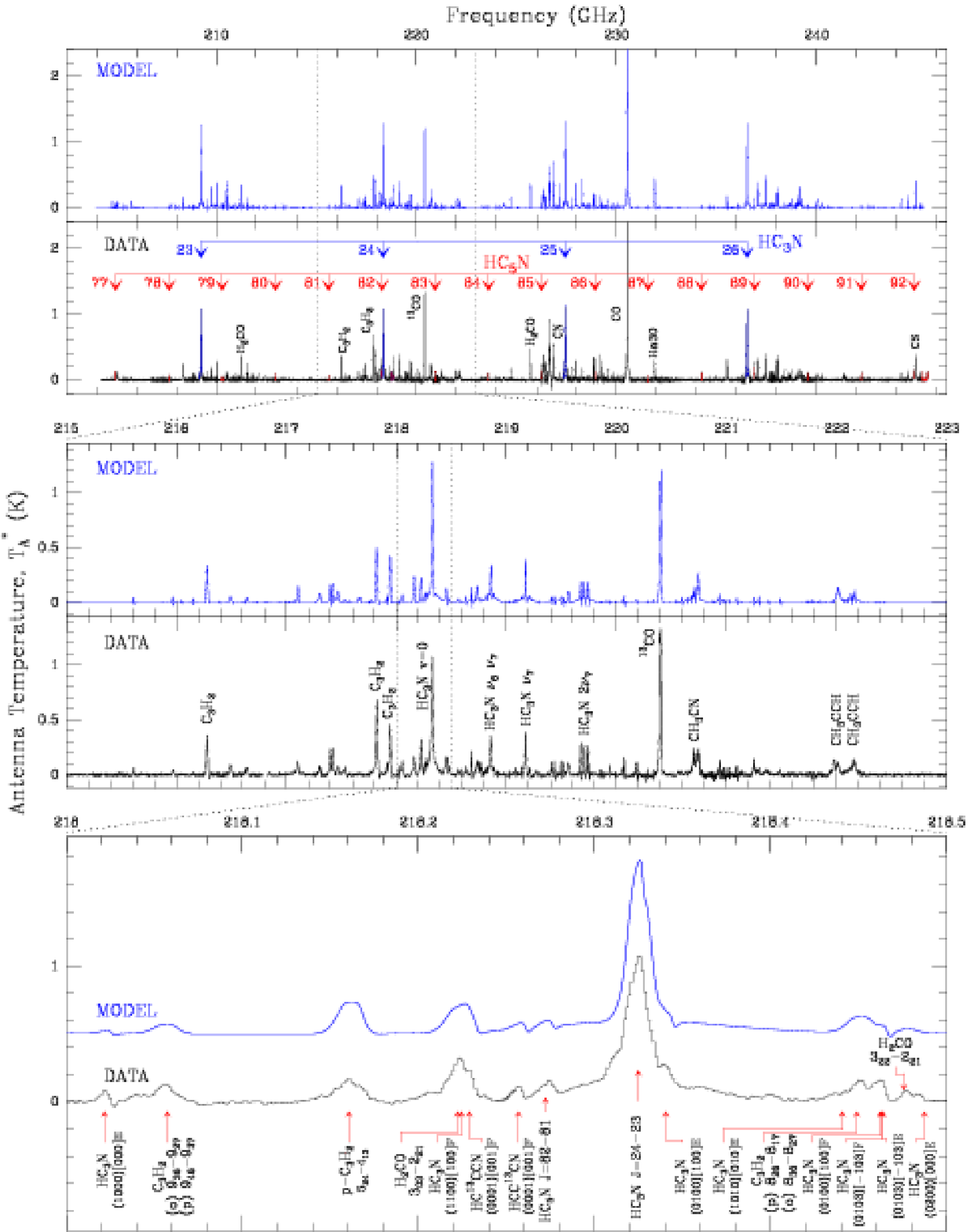}
 \caption{Same as previous figure but for the frequency 
range 204.5-245.0 GHz (lower part of the 1.3 mm window).}
              \label{Fig0c}%
    \end{figure*}

\begin{figure*}[h]
 \centering
 \includegraphics[angle=0,width=.8\textwidth]{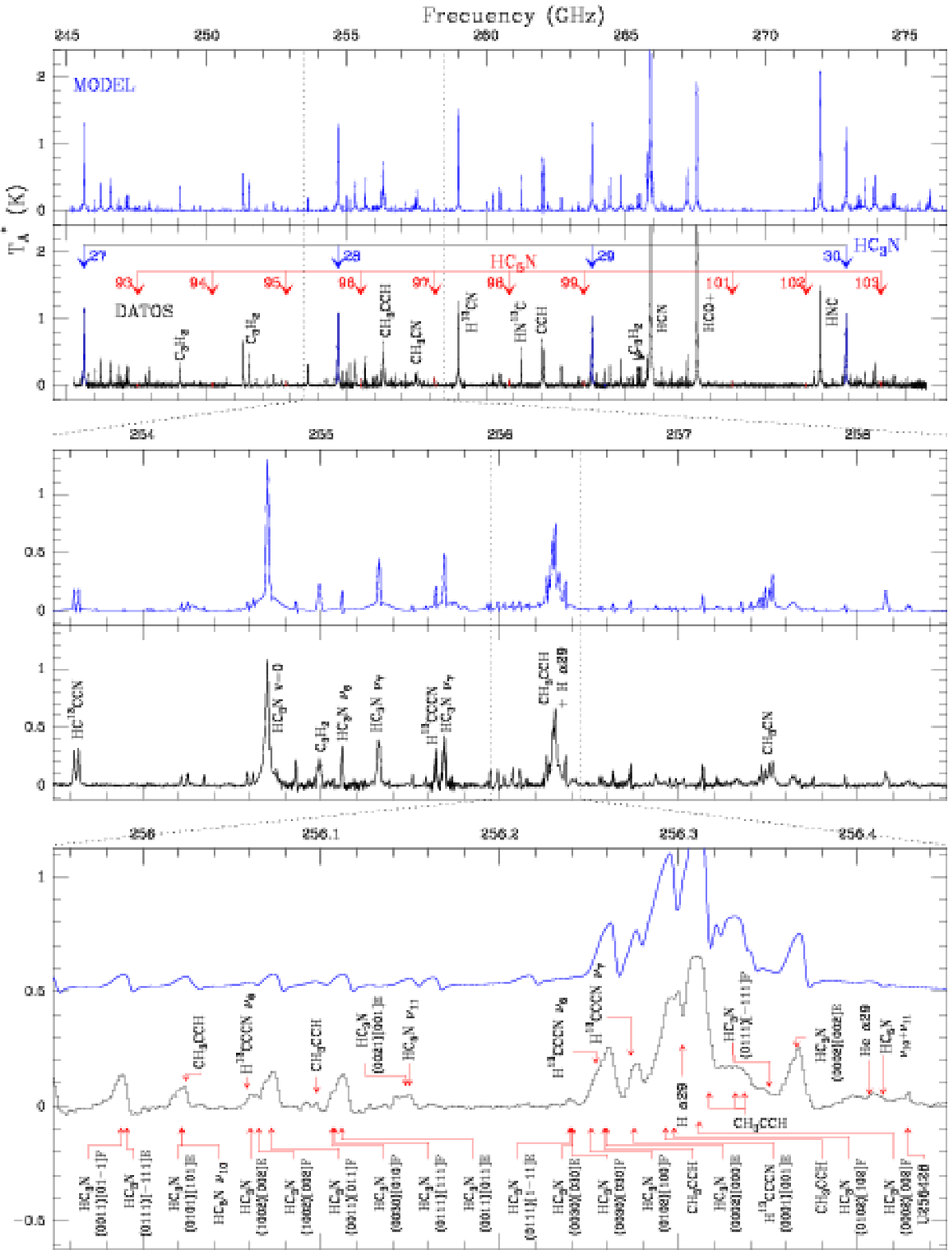}
 \caption{Same as previous figure but for the frequency  
range 245.0-275.5 GHz (upper part of the 1.3 mm window).}
              \label{Fig0d}%
    \end{figure*}

The overall results of the line survey presented in figures 
\ref{Fig0a}, \ref{Fig0b}, \ref{Fig0c}, and \ref{Fig0d} show 
that the most intense lines (above 0.2 K in 
T$_{A}^{*}$) belong to a very limited number of species (H recombination lines, 
C$_{3}$H$_{2}$, HCN, HCO+, CN, CO, CS, H$_{2}$CO, CH$_{3}$CN, CCH and HC$_{3}$N). 
They also show how, as frequency increases,  
the absorption part of the line profile becomes less important with respect to the 
emission part in species showing P-Cygni profiles, indicating 
that the spectral behavior of the continuum flux plays a fundamental role in the evolution 
of the absorption/emission ratio. The absolute energy levels of the transitions play also 
a role, but less important. The clustering of lines belonging to HC$_{3}$N and its 
isotopic and vibrationally excited species every $\sim$9 GHz is also nicely delineated. 
The detailed zoom (500 MHz) at the bottom of each figure illustrates the line identification process. Individual 
figures at this level of detail, including a comparison with the model, have been created for the whole data 
set and are available only as on-line material. Some spectra are so crowded with lines that zooms 
showing only 250 MHz of data have been necessary. A set of tables with the format of Table \ref{id-lines}, only 
available also as 
on-line material, provide the rest frequencies of all identified features and the complete 
quantum numbers of the transition (usually omitted in the figures due to lack of space). The 
labeling is quite straightforward for all molecules with the exception of HC$_3$N that, we should 
remind, is responsible for about 55\% of the features seen in the survey. For this molecule the 
lines are labeled by the rotational quantum numbers (J$_{up}$--J$_{low}$), the vibrational 
quantum numbers of the four (out of seven) lowest vibrational modes ($v_4$,$v_5$,$v_6$,$v_7$) and   
the $\ell$-doubling parameters for the bending modes: [$\ell_{5}$ $\ell_{6}$ $\ell_{7}$ ]. The 
complete set of frequencies for this molecule has been obtained from Fayt et al. (2004), and 
the agreement is excellent. For 
HC$_5$N we use a similar notation with quantum numbers of the three (out of eleven) lowest vibrational modes 
($v_9$,$v_{10}$,$v_{11}$) and the necessary $\ell$-doubling parameters.

\subsection{Line identification} 

The line identification process has been based on J. Cernicharo's own molecular 
rotational line catalog. The catalog contains now $\sim$1300 molecules, radicals and atoms, 
and the number of lines in the frequency range 80-280 GHz exceeds 600000. 

Although part of the line identification could be done by means of an automated 
procedure (mainly designed to recognize the lines from the cyanopolyynes 
HC$_{3}$N and HC$_{5}$N), the spectra are sometimes so crowded with lines that an important 
part of the identification and labeling work has to be done manually, specially  
taking into account that the greatest interest is usually in the weak features. 

The molecular species with identified lines in the millimeter spectrum of CRL 618 are listed in 
Tables \ref{table-species} and \ref{table-species-bis}. Very few are O-bearing molecules (only SiO, H$_2$CO 
and HCO$^+$ in addition to CO), and there is also CS, and recombination lines 
of Hydrogen and Helium. The most prominent features in the spectrum are due to the 
cyanopolyynes family, mainly HC$_{3}$N, for which we have identified lines 
from more than 30 isotopic, isomeric or vibrationally excited forms. The key for deducing 
the basic physical parameters of CRL 618 has been to count on several hundred of lines of 
HC$_{3}$N with a wide range of excitation conditions to probe the different gas regions of 
this source (see P04 and P05).

We know nevertheless that this object has to have some relatively large carbon chains, based 
on the discoveries published by Cernicharo et al. (2001a,b). Compared to 
other cases such as Orion or IRC+10216, the number of U-lines in the CRL 618 survey is very 
small: Only 74 or one per $\sim$2.1 GHz as an average in the surveyed range. Although 
several species are candidates to explain a few of these U lines, their complex rotational 
spectrum should display other features arising above the noise level that are not seen, and 
therefore the assignment could not be established. Table \ref{u-lines} gives the frequencies 
and peak emission and absorption (if present) values for the unidentified features after 
subtracting the continuum level. The search for some particular species is described in 
section \ref{sct:search}. 

Detailed tables and figures showing the $\sim$3100 features that could be identified with the 
help of the model described in section  \ref{sct:model} are available as on-line material only.

\subsection{Model} 
\label{sct:model}

A model aimed at reproducing the whole millimeter wave spectrum of CRL 618 as accurately as possible with the minimum 
number of parameters has been developed. To achieve its completion, four basic steps have been 
followed:

\begin{enumerate}
\item Model for the continuum flux considering a fixed size of the region from which it emerges, 
its brightness temperature at a given frequency, and a spectral index. The values used, 0.27'' 
size, 3900 K at 200 GHz and $(\nu/\nu_0)^{-1.12}$ behavior, were found and discussed in P04.

\item Model for the inner (1.5'') slowly expanding molecular envelope ({\it SEE}) using lines from vibrationally 
excited states of HC$_3$N (the best tool for this task).  The average temperature of the region 
(250-275 K), HC$_3$N column density in front of the continuum source ($\sim$3$\cdot$10$^{17}$ cm$^{-2}$),  
and details about the morphology and the velocity field, were also obtained in P04.

\item Model for the high velocity wind ({\it HVW}), reaching $\sim$200 kms$^{-1}$, presumably quite collimated, 
noticeable in some abundant species such as CO, HCN and $v$=0 HC$_3$N (see P05).

\item Model for the colder and outer, with respect to the {\it SEE}, cold circumstellar shell ({\it CCS}) that is responsible 
for most of the rotational emission from  $\nu_7$ and $v$=0 HC$_3$N, and of $v$=0 HC$_5$N. This gas component 
is also dominant in the emission of several other species. The average temperature ($\sim$30 K), position (3.0'' 
to 4.5'' diameter), velocity field, density distribution, and abundance ratio of all species with respect to 
HC$_3$N, have been obtained in P05 and PC06. 
\end{enumerate}


The model goes into some detail on the morphology of the object. For example, the geometry is not 
spherical. The {\it SEE} has an elongated shape, the {\it HVW} is collimated, and the {\it CCS} has a bipolar density 
distribution. The velocity field has both radial and azimuthal components. The azimuthal symmetry is 
broken  because the line of sight is inclined with respect to the symmetry axis. See Fig. 4 of P04 
and Fig. 3 of P05 for more details. As a result, as discussed in section 3.2 of PC06, quadratures 
in 4 different variables are necessary for the calculations (frequency, impact parameter, path along 
the line of sight, and angle in the plane perpendicular to the line of sight). All species listed 
in Table \ref{table-species} have been included in the general calculations (except stated otherwise). 

In P04, P05 and PC06, the calculations were performed for a selection of lines of each species that met 
several requirements such as avoiding line blendings and representing well the energy range probed by 
each species. This allowed us to draw a quite precise picture of the structure, physical conditions and chemical abundances 
in the source with reasonable computation times. Now, the model has been run for all surveyed frequencies in 
order to check its consistency and to be provided as a final product of this study together with the 
whole and labeled data set.

\subsection{Search for Trace Molecules}
\label{sct:search}
   
As the complexity of the molecules increases, their abundances are usually lower and their 
partition functions larger. In order to identify a new molecule in the data, it is necessary 
to run a model for a large set of transitions. Our search for new species has been based on 
possible products from simple reactions using molecules and radicals already identified in this 
survey and in IR observations. 


\subsubsection{C$_6$H radical}
The abundance of C$_2$H and CN in CRL 618 allows an efficient growth of 
HC$_{2n+1}$N and C$_{2n}$H$_2$ molecules (see the detection of the polyyne chains C$_2$H$_2$, 
C$_4$H$_2$, and C$_6$H$_2$ in Cernicharo et al. 2001; and the polymerization 
of HCN in Cernicharo 2004 and P05). As a consequence, the linear radicals C$_{2n}$H, with 
only one free bond, are much more abundant in this object than C$_{2n+1}$H, with three free 
bonds. Therefore, 
we expected to detect the C$_6$H linear radical. However, the large rotational partition function of 
this species combined with a decrease in abundance with respect to C$_{4}$H and also the   
the higher energy levels of C$_6$H that are sounded above 80 GHz (well above the Boltzmann peak corresponding to a temperature of 
$\sim$30 K) with respect to the lighter radicals, make in fact its detectability marginal at the sensitivity level of the survey. 
 
\subsubsection{C$_5$N radical}
Although the radical C$_3$N is more abundant than the combination of c-C$_3$H and l-C$_3$H, C$_5$N is 
not detected in the survey, contrary to C$_5$H. The latter is the less abundant main species detected 
in the survey. Assuming a similar abundance for C$_5$N its non detection could be explained by its 
lower dipole moment (3.385 Debyes instead of 4.881 Debyes for C$_5$H). In any case, above 80 GHz the 
rotational lines of both molecules are past the Boltzmann peak corresponding to the temperature 
of the cold circumstellar shell of CRL 618.

\subsubsection{H$_2$C$_4$ and H$_2$C$_3$ radicals}
Double C bounds seem to be rare in CRL 618 compared with carbon chains alternating single and triple 
bounds (as in the cyanopolyynes, polyynes and related radicals). This is further confirmed by our search 
of the H$_2$C$_4$ and H$_2$C$_3$ radicals (in which all carbons are doubly bounded) that has resulted on 
non detection for H$_2$C$_4$ and a marginal one for  H$_2$C$_3$. Although some features in the spectra 
are located at H$_2$C$_4$ and H$_2$C$_3$ 
frequencies, a calculation of the whole spectrum to reproduce those features gives inconsistencies at many 
other frequencies (non-detections where a line should be expected).

\subsubsection{CH$_2$CHCN}
This is so far the only positive new finding. This molecule has $\mu_a$=3.815 Debyes and $\mu_b$=0.894 Debyes 
transitions that result on many weak features in our 
survey (see Fig. \ref{Fig0b}, bottom panel). 
We have found an abundance $\sim$15 times less than 
that of HC$_3$N in the {\it CCS}. This value is intermediate between those found for CH$_3$CN 
and CH$_3$CCH, although there is evidence that part of the signal detected from the last 
two molecules comes from the {\it SEE} (see PC06). The relatively similar abundances found for 
the three species tell us about a similar chemistry leading to their formation; see 
Cernicharo (2004) for details about chemical routes in CRL 618.

\subsubsection{C$_6$H$_5$CN and C$_6$H$_5$CCH}
Given the evidence in support of the detection of benzene (C$_6$H$_6$) and the large abundances of the radicals CN and CCH in CRL 618, we 
have checked for the presence of these two molecules but unfortunately with negative results. The large partition function 
of these species combined with a column density of benzene of 2$\times$10$^{15}$ cm$^{-2}$ at most in this object 
(Cernicharo et al. 2001) in fact imply that the intensity of the lines cannot be expected to arise above the    
sensitivity level of the survey.  

\subsubsection{CH$_2$CN}
This radical, that in principle could be expected to be detectable as it should be formed in 
similar routes to those of the 
abundant CH$_3$CN in CRL 618 has been the subject of an specific search. Only 
some features around 161.95 GHz and 241.35 GHz (see table \ref{u-lines}) seem to be related to this 
radical, but the detection cannot be confirmed because there is no correspondence at other 
frequencies where a line should be expected.

\subsubsection{$^{13}$C substitutions of c-C$_3$H$_2$}
Starting from the abundance of c-C$_3$H$_2$ and considering a ratio $^{12}$C/$^{13}$C of 40 in the 
{\it CCS}, found in P05 and PC06, it would result that H$^{13}$CCCH at least could be detectable. Its non detection indicates 
that perhaps the value 40 should be understood as a lower limit. Further evidence for this is provided 
by H$_2^{13}$CO, $^{13}$CS, $^{13}$CCH, and C$^{13}$CH, that are not seen either. We should remind that the ratio  
$^{12}$C/$^{13}$C is only about 15 in the inner {\it SEE}, possibly due to injection of $^{13}$C-rich material generated 
in a late thermal pulse to the inner parts of the gas outflows of CRL 618. The result presented in this section, therefore, 
further strengths our (PC06) finding of quite different $^{12}$C/$^{13}$C abundance ratios in the {\it SEE} and {\it CCS} regions of CRL 618.

\subsection{Comparison with PPN chemical models}
\label{sct:modelcomparison}
The study of Woods et al. (2003) provides chemical abundances in a region within 8.9$\cdot$10$^{15}$ cm from the central star, 
which roughly coincides with the SEE of our model. Their predictions provide small [HC$_3$N]/[HC$_5$N]/[HC$_7$N] ratios, just as 
we observed, that support the efficiency of polimerization in creating polyyne and cyanoplyyne chains in this inner SEE (already 
reported in P05 and in agreement with predictions in Cernicharo 2004). The predictions made in Woods et al. (2003) for radicals 
such as CN, CCH, C$_4$H, C$_3$N,... show relative 
abundances with respect to HC$_3$N of less than 10$^{-2}$ (except $\sim$10$^{-1}$ for C$_3$N) which would make them very difficult 
to detect in the SEE. This is confirmed by our data because the emission we detect from those radicals is 
dominated by the CCS component (almost no signs of absorption). The reason for this is that at the densities and temperatures of 
the SEE, reactions with H$_2$ would be very efficient to transfer those radicals into polyyne or cyanopolyyne species (that dominate 
in this region) whereas in the much colder CCS (T$\sim$30 K) the reactions with H$_2$ would barely operate.

\section{Conclusions}

This paper has presented the final products of the most complete spectroscopic study at millimeter wavelengths 
of a protoplanetary nebula carried out up to date. First, the whole data set with line identifications for about 3100 features 
has been presented. 
The number of remaining unidentified features is unprecedentedly low, only one per $\sim$2.1 GHz (a total of 74 in 155 GHz). 
Second, the whole data set has been compared with a complete model generated from the basic physical conditions and chemical 
abundances deduced in the previous papers of this series (P04, P05, and PC06). Attempts have been made to search for new 
molecular species, not published in those works, but so far the only new finding is CH$_2$CHCN in the outer Cold Circumstellar 
Shell of CRL 618, with a chemical formation route that must be similar to those of CH$_3$CN and CH$_3$CCH. The resulting 
morphological, physical and chemical picture of CRL 618 that has emanated from this study is unprecendently precise thanks 
to the wide range of energy levels in the detected  molecular lines that have subsequently been analyzed to probe the object. We 
have been able to provide details much beyond the angular resolution of the observations. Only interferometric observations could 
have provided a similar level of detail. However present-day interferometry could have concentrated on only a few lines and so a lot of 
information on the chemical composition would be missed. The present model 
can now be easily extended to the frequency range 280-360 GHz, for which another survey is almost completed with the Caltech 
Submillimeter Observatory. The Herschel Space Observatory will provide the opportunity to access higher frequencies into the 
submillimeter range with an unprecedented sensitivity and limited dilution compared with ISO. This will allow to access new 
molecular species and to complete the study of the continuum emission in this source. The advent of EVLA and ALMA will allow 
direct mapping of the structures delineated in this work and the possibility of a nearly continuous coverage from 300 MHz 
to 950 GHz. The large collecting area should allow to extend our view of the chemical system.

\acknowledgments
We acknowledge the support of the IRAM-30m staff during the long completion of the 
line survey. This work has also been supported by Spanish DGES and PNIE 
grants ESP2002-01627, AYA2002-10113-E and AYA2003-02785-E.

\appendix
\section{Online materials}

Tables A1, A2 and A3 contain the identified features in the CRL 618 millimeter 
wave survey in the following frequency ranges: 80.3-115.5 GHz, 131.2-179.1 GHz, and
the 204.5-276.0 GHz.  Also included are 
online only figures showing the millimeter spectra.

\begin{deluxetable}{llllll}
\tablecolumns{6} 
\tablewidth{0pc} 
\tablecaption{Identified features in the 80.3-115.5 GHz range in the CRL 618 survey presented in this paper.}
\tablehead{ 
\colhead{$\nu_0$(MHz)} & \colhead{Molecule} & \colhead{Transition/$v$-state} & \colhead{$\nu_0$(MHz)} & \colhead{Molecule} & \colhead{Transition/$v$-state}  }
\startdata
   80305.928 & HC$_{5}$N                & J=30-29 (003)[-1] &
   80380.000 & HC$_{5}$N                & J=30-29 (003)[-3]                      \\
   80388.107 & l-C$_{3}$H  $^{2}\Pi_{3/2}$   &  (J,F)=(7/2,4)-(5/2,3) a &
   80388.442 & l-C$_{3}$H  $^{2}\Pi_{3/2}$   &  (J,F)=(7/2,3)-(5/2,2) a \\
   80420.646 & l-C$_{3}$H  $^{2}\Pi_{3/2}$   &  (J,F)=(7/2,4)-(5/2,3) b &
   80422.052 & l-C$_{3}$H  $^{2}\Pi_{3/2}$   &  (J,F)=(7/2,3)-(5/2,2) b \\
   80446.196 & HC$_{5}$N                & J=30-29 (003)[1]     &
   80559.000 & U080559                        &  \\
   80618.000 & U080618                        &  &
   80723.188 & C$_{3}$H$_{2}$ para    & J$_{K+,K-}$=4$_{2,2}$-4$_{1,3}$ \\
\enddata
\tablecomments{Table A1 is published in its entirety in the electronic 
edition of the {\it Astrophysical Journal}.  A portion is shown here 
for guidance regarding its form and content.}
\end{deluxetable}

\begin{deluxetable}{llllll}
\tablecolumns{6} 
\tablewidth{0pc} 
\tablecaption{Identified features in the 131.2-179.1 GHz range in the CRL 618 survey presented in this paper.}
\tablehead{ 
\colhead{$\nu_0$(MHz)} & \colhead{Molecule} & \colhead{Transition/$v$-state} & \colhead{$\nu_0$(MHz)} & \colhead{Molecule} & \colhead{Transition/$v$-state}  }
\startdata
  131256.830   & MgNC                             & J=23/2-21/2 u  &
  131267.422   & CH$_{2}$CHCN               & J$_{K+,K-}$ = 14$_{0,14}$-13$_{0,13}$ \\
  131277.716   & C$_{5}$H $^{2}\Pi_{1/2}$   & J=55/2-53/2 a &
  131285.894   & C$_{5}$H $^{2}\Pi_{1/2}$   & J=55/2-53/2 b \\
  131283.629   & HC$_{5}$N                  & J=49-48 (003)[-3]   &
  131286.219   & HC$_{5}$N                  & J=49-48 (003)[+3]   \\
  131376.100   & HC$_{5}$N                  & J=49-48 (003)[1]                      &
  132179.132   & C$_{4}$H                   & J=14-13 $\nu_{7}$ \\
  132246.366   & H$^{13}$CCCN    & J=15-14 &
  132381.516   & U132378                          &  \\
\enddata
\tablecomments{Table A2 is published in its entirety in the electronic 
edition of the {\it Astrophysical Journal}.  A portion is shown here 
for guidance regarding its form and content.}
\end{deluxetable}

\begin{deluxetable}{llllll}
\tablecolumns{6} 
\tablewidth{0pc} 
\tablecaption{Identified features in the 204.5-276.0 GHz range in the CRL 618 survey presented in this paper.}
\tablehead{ 
\colhead{$\nu_0$(MHz)} & \colhead{Molecule} & \colhead{Transition/$v$-state} & \colhead{$\nu_0$(MHz)} & \colhead{Molecule} & \colhead{Transition/$v$-state}  }
\startdata
  204788.938    & c-C$_{3}$H$_{2}$            & J$_{K+,K-}$ = 2$_{2,0}$-3$_{1,0}$    &
  204970.268    & HC$_{5}$N                   & J=77-76  \\
  205003.000    & U205003                     &    &
  205018.109    & CH$_{3}$CCH                 & J=12-11 K=4   \\
  205045.497    & CH$_{3}$CCH                 & J=12-11 K=3    &
  205065.067    & CH$_{3}$CCH                 & J=12-11 K=2  \\
  205076.813    & CH$_{3}$CCH                 & J=12-11 K=1    &
  205307.362    & HC$_{5}$N                   & J=77-76 (001)[-1]   \\
  205308.360    & HC$_{5}$N                   & J=77-76 (010)[-1]     &
  205385.073    & HC$_{5}$N                   & J=77-76 (010)[+1]   \\
\enddata
\tablecomments{Table A3 is published in its entirety in the electronic 
edition of the {\it Astrophysical Journal}.  A portion is shown here 
for guidance regarding its form and content.}
\end{deluxetable}

{}

\begin{deluxetable}{lccccccc}
\tablecolumns{8} 
\tablewidth{0pc} 
\tablecaption{Cyanopolyyne species included in the model and J$_{up}$ and/or E$_{up}$ range in which they 
appear in the data. We separate isotopic substitutions and vibrationally excited states 
(their vibrational energy is also given) 
of those species for which several of them have been detected.}
\tablehead{ 
\colhead{\footnotesize Molecule and} & \colhead{\footnotesize  E$_{vib}$}   & \colhead{\footnotesize  J$_{up}$} & \colhead{\footnotesize   Range E$_{rot}$} & 
\colhead{\footnotesize  Total Num.} & \colhead{\footnotesize   [X]/[HC$_3$N]} & \colhead{\footnotesize   [X]/[HC$_3$N]} & \colhead{\footnotesize   [X]/[HC$_3$N]} \\
\colhead{vib. state} & \colhead{\footnotesize  (cm$^{-1}$)} & \colhead{\footnotesize  range} & \colhead{\footnotesize    (cm$^{-1}$)} & \colhead{\footnotesize  of labels}  & 
\colhead{\footnotesize  in {\it SEE}} & \colhead{\footnotesize  in {\it HVW}} & \colhead{\footnotesize  in {\it CCS} }   }
\startdata 
\footnotesize HC$_3$N (0000) &\footnotesize    &\footnotesize  9-30 &\footnotesize  15-205 &\footnotesize  17 &\footnotesize  0.4092 &\footnotesize  0.2325 &\footnotesize  0.9905 \\
\footnotesize HC$_3$N (0001) &\footnotesize  223  &\footnotesize  9-30 &\footnotesize  15-205 &\footnotesize  34 &\footnotesize  0.2412 &\footnotesize  0.1859 &\footnotesize  0.0095 \\
\footnotesize HC$_3$N (0002) &\footnotesize  446  &\footnotesize  9-30 &\footnotesize  15-205&\footnotesize  51 &\footnotesize  0.1065 &\footnotesize  0.1115 &\footnotesize  - \\
\footnotesize HC$_3$N (0010) &\footnotesize  499  &\footnotesize  9-30 &\footnotesize  15-205 &\footnotesize   34 &\footnotesize  0.0531 &\footnotesize  0.0598 &\footnotesize  - \\
\footnotesize HC$_3$N (0100) &\footnotesize  663  &\footnotesize  9-30 &\footnotesize  15-205 &\footnotesize  34 &\footnotesize  0.0216 &\footnotesize  0.0304 &\footnotesize  - \\
\footnotesize HC$_3$N (0003) &\footnotesize  669  &\footnotesize  9-30 &\footnotesize  15-205 &\footnotesize   68 &\footnotesize  0.0418 &\footnotesize  0.0594 &\footnotesize  - \\
\footnotesize HC$_3$N (0011) &\footnotesize  721  &\footnotesize  9-30 &\footnotesize  15-205 &\footnotesize  68  &\footnotesize  0.0313 &\footnotesize  - &\footnotesize  - \\
\footnotesize HC$_3$N (1000) &\footnotesize  880  &\footnotesize  9-30 &\footnotesize  15-205 &\footnotesize  17 &\footnotesize   0.0033 &\footnotesize  - &\footnotesize  - \\
\footnotesize HC$_3$N (0101) &\footnotesize  886  &\footnotesize  9-30 &\footnotesize  15-205 &\footnotesize   68 &\footnotesize  0.0127 &\footnotesize  - &\footnotesize  - \\
\footnotesize HC$_3$N (0004) &\footnotesize  892  &\footnotesize  9-30  &\footnotesize  15-205 &\footnotesize  77  &\footnotesize  0.0154 &\footnotesize  - &\footnotesize  - \\
\footnotesize HC$_3$N (0012) &\footnotesize  944  &\footnotesize  9-30  &\footnotesize  15-205 &\footnotesize  100 &\footnotesize  0.0138 &\footnotesize  - &\footnotesize  - \\
\footnotesize HC$_3$N (0020) &\footnotesize  998  &\footnotesize  9-30 &\footnotesize  15-205 &\footnotesize  49 &\footnotesize  0.0052  &\footnotesize  - &\footnotesize  - \\
\footnotesize HC$_3$N (1001) &\footnotesize  1103 &\footnotesize  11-30 &\footnotesize 24-205 &\footnotesize  27 &\footnotesize  0.0019 &\footnotesize  - &\footnotesize  - \\
\footnotesize HC$_3$N (0102) &\footnotesize  1109 &\footnotesize  11-30 &\footnotesize 24-205 &\footnotesize  88 &\footnotesize  0.0056 &\footnotesize  - &\footnotesize  - \\
\footnotesize HC$_3$N (0005) &\footnotesize  1115 &\footnotesize  11-30 &\footnotesize 24-205 &\footnotesize   66 &\footnotesize  0.0054 &\footnotesize  - &\footnotesize  - \\
\footnotesize HC$_3$N (0110) &\footnotesize  1162 &\footnotesize  11-30 &\footnotesize 24-205 &\footnotesize   62 &\footnotesize  0.0029 &\footnotesize  - &\footnotesize  - \\
\footnotesize HC$_3$N (0013) &\footnotesize  1168 &\footnotesize  12-30 &\footnotesize 29-205 &\footnotesize   98 &\footnotesize 0.0055  &\footnotesize  - &\footnotesize  - \\
\footnotesize HC$_3$N (0021) &\footnotesize  1221 &\footnotesize  15-30 &\footnotesize  46-205 &\footnotesize   64 &\footnotesize 0.0031  &\footnotesize  - &\footnotesize  - \\
\footnotesize HC$_3$N (1002)   &\footnotesize   1326 &\footnotesize  15-30  &\footnotesize  46-205 &\footnotesize  26  &\footnotesize  0.0009  &\footnotesize  -  &\footnotesize  - \\
\footnotesize HC$_3$N (0103)   &\footnotesize   1332 &\footnotesize  15-30  &\footnotesize  46-205  &\footnotesize  90  &\footnotesize  0.0023  &\footnotesize  -  &\footnotesize  - \\
\footnotesize HC$_3$N (0006)           &\footnotesize   1338 &\footnotesize  15-30  &\footnotesize  46-205 &\footnotesize  22  &\footnotesize  0.0019  &\footnotesize  -  &\footnotesize  - \\
\footnotesize HC$_3$N (1010)    &\footnotesize  1379  &\footnotesize  16-30     &\footnotesize  53-205 &\footnotesize  9  &\footnotesize  0.0004 &\footnotesize  -  &\footnotesize  - \\
\footnotesize HC$_3$N (0111) &\footnotesize  1385 &\footnotesize  17-30 &\footnotesize  60-205 &\footnotesize  57 &\footnotesize  0.0017  &\footnotesize   - &\footnotesize    -\\
\footnotesize HC$_3$N (0014)  &\footnotesize  1431   &\footnotesize  17-30 &\footnotesize  60-205 &\footnotesize  56  &\footnotesize  0.0016  &\footnotesize   - &\footnotesize  - \\
\footnotesize HC$_3$N (0022)  &\footnotesize  1464  &\footnotesize  19-30 &\footnotesize  75-205 &\footnotesize  55  &\footnotesize  0.0012  &\footnotesize   - &\footnotesize  - \\
\footnotesize HC$_3$N (0030)           &\footnotesize  1497  &\footnotesize  23-30 &\footnotesize  111-205 &\footnotesize  20  &\footnotesize  0.0005  &\footnotesize  -  &\footnotesize  - \\
\footnotesize HC$_3$N (1100)    &\footnotesize  1543  &\footnotesize  23-27 &\footnotesize  111-165 &\footnotesize  5  &\footnotesize  0.0002 &\footnotesize  -  &\footnotesize  - \\
\footnotesize HC$_3$N (0201)    &\footnotesize  1549  &\footnotesize 23-30  &\footnotesize  111-205 &\footnotesize  19  &\footnotesize 0.0005   &\footnotesize  -  &\footnotesize  - \\
\footnotesize HC$_3$N (1003)    &\footnotesize  1549  &\footnotesize 23-29  &\footnotesize  111-191 &\footnotesize  14  &\footnotesize  0.0003  &\footnotesize  -  &\footnotesize  - \\
\footnotesize HC$_3$N (0104)    &\footnotesize  1555  &\footnotesize 23-29    &\footnotesize  111-191 &\footnotesize  21  &\footnotesize 0.0008  &\footnotesize  -  &\footnotesize  - \\
\footnotesize HC$_3$N (1011)    &\footnotesize  1602  &\footnotesize 24-30  &\footnotesize  121-205 &\footnotesize  8  &\footnotesize 0.0003   &\footnotesize  -  &\footnotesize  - \\
\footnotesize HC$_3$N (0112)    &\footnotesize  1608  &\footnotesize 26-30    &\footnotesize  142-205 &\footnotesize  21  &\footnotesize  0.0007  &\footnotesize  -  &\footnotesize  - \\
\footnotesize HC$_3$N (0015)    &\footnotesize  1608  &\footnotesize   26    &\footnotesize  142-154 &\footnotesize  1  &\footnotesize  0.0007  &\footnotesize  -  &\footnotesize  - \\
\footnotesize H$^{13}$CCCN (0000) &\footnotesize     &\footnotesize  10-31 &\footnotesize 20-218 &\footnotesize  18  &\footnotesize  0.0273 &\footnotesize  0.0155 &\footnotesize  0.0248 \\
\footnotesize HC$^{13}$CCN (0000) &\footnotesize     &\footnotesize  9-30  &\footnotesize 15-205  &\footnotesize  17 &\footnotesize  0.0273 &\footnotesize  0.0155 &\footnotesize  0.0248 \\
\footnotesize HCC$^{13}$CN (0000) &\footnotesize     &\footnotesize  9-30  &\footnotesize 15-205 &\footnotesize  17  &\footnotesize  0.0273 &\footnotesize  0.0155 &\footnotesize  0.0248 \\
\footnotesize H$^{13}$CCCN (0001) &\footnotesize   223  &\footnotesize  10-31  &\footnotesize 20-218 &\footnotesize  36 &\footnotesize  0.0161 &\footnotesize  - &\footnotesize  - \\
\footnotesize HC$^{13}$CCN (0001) &\footnotesize  223   &\footnotesize  9-30  &\footnotesize 15-205 &\footnotesize  34 &\footnotesize  0.0161 &\footnotesize  - &\footnotesize  - \\
\footnotesize HCC$^{13}$CN (0001) &\footnotesize  223   &\footnotesize  9-30  &\footnotesize 15-205 &\footnotesize  35 &\footnotesize  0.0161 &\footnotesize  - &\footnotesize  - \\
\footnotesize H$^{13}$CCCN (0002) &\footnotesize  446  &\footnotesize  16-30  &\footnotesize 53-205  &\footnotesize  8 &\footnotesize 0.0071  &\footnotesize  - &\footnotesize  - \\
\footnotesize HC$^{13}$CCN (0002) &\footnotesize  446   &\footnotesize 17-26   &\footnotesize 60-154  &\footnotesize  9 &\footnotesize 0.0071  &\footnotesize  - &\footnotesize  - \\
\footnotesize HCC$^{13}$CN (0002) &\footnotesize  446   &\footnotesize 16-28   &\footnotesize 53-178 &\footnotesize  13 &\footnotesize  0.0071 &\footnotesize  - &\footnotesize  - \\
\footnotesize H$^{13}$CCCN (0010) &\footnotesize   499  &\footnotesize  12-30  &\footnotesize 29-205  &\footnotesize  26 &\footnotesize  0.0035 &\footnotesize  - &\footnotesize  - \\
\footnotesize HC$^{13}$CCN (0010) &\footnotesize  499   &\footnotesize  10-30 &\footnotesize 20-205 &\footnotesize  32  &\footnotesize  0.0035 &\footnotesize  -&\footnotesize  - \\
\footnotesize HCC$^{13}$CN (0010) &\footnotesize  499   &\footnotesize  10-30  &\footnotesize 20-205 &\footnotesize   29 &\footnotesize  0.0035 &\footnotesize  - &\footnotesize  - \\
\footnotesize H$^{13}$CCCN (0003) &\footnotesize   669  &\footnotesize    &\footnotesize  &\footnotesize  1 &\footnotesize 0.0028  &\footnotesize  - &\footnotesize  - \\
\footnotesize HCCNC &\footnotesize    &\footnotesize  10-29 &\footnotesize   &\footnotesize  15 &\footnotesize  20-191 &\footnotesize  - &\footnotesize  0.0250 \\
\hline
\footnotesize TOTAL  HC$_3$N & & & & 1736 & & & \\
\hline
\footnotesize HC$_5$N $v$=0 &\footnotesize    &\footnotesize  31-104 &\footnotesize  39-485 &\footnotesize  55 &\footnotesize  0.0132 &\footnotesize  - &\footnotesize  0.2794 \\  
\footnotesize HC$_5$N $\nu_{11}$ &\footnotesize  105 &\footnotesize  31-104 &\footnotesize  39-485 &\footnotesize  118  &\footnotesize  0.0149 &\footnotesize  - &\footnotesize  0.0451 \\ 
\footnotesize HC$_5$N 2$\nu_{11}$ &\footnotesize  210  &\footnotesize  31-104 &\footnotesize  39-485 &\footnotesize  147 &\footnotesize  0.0126 &\footnotesize   - &\footnotesize  0.0055 \\ 
\footnotesize HC$_5$N $\nu_{10}$ &\footnotesize  230 &\footnotesize  31-104 &\footnotesize  39-485 &\footnotesize  105 &\footnotesize   0.0075 &\footnotesize  - &\footnotesize  0.0023 \\ 
\footnotesize HC$_5$N 3$\nu_{11}$ &\footnotesize  315  &\footnotesize  31-95 &\footnotesize  39-402  &\footnotesize  152 &\footnotesize   0.0094 &\footnotesize  -  &\footnotesize  0.0006 \cr
\footnotesize HC$_7$N $v$=0 &\footnotesize    &\footnotesize  72-102 &\footnotesize 102-198  &\footnotesize 26 &\footnotesize -  &\footnotesize  - &\footnotesize  0.0503 \cr
\footnotesize HC$_7$N $\nu_{15}$ &\footnotesize  62  &\footnotesize  79 &\footnotesize 113-116  &\footnotesize  1  &\footnotesize -  &\footnotesize  - &\footnotesize  0.0052 \cr  
\hline
\footnotesize TOTAL  HC$_{2n+1}$N & & & & 2340 & & & \cr
\enddata
\label{table-species}
\end{deluxetable}

\begin{deluxetable}{lccccccc}
\tablecolumns{8} 
\tablewidth{0pc} 
\tablecaption{Other molecules included in the model, presented as in Table  \ref{table-species}.}
\tablehead{ 
\colhead{\footnotesize Molecule and} & \colhead{\footnotesize  E$_{vib}$}   & \colhead{\footnotesize  J$_{up}$} & \colhead{\footnotesize   Range E$_{rot}$} & 
\colhead{\footnotesize  Total Num.} & \colhead{\footnotesize   [X]/[HC$_3$N]} & \colhead{\footnotesize   [X]/[HC$_3$N]} & \colhead{\footnotesize   [X]/[HC$_3$N]} \\
\colhead{vib. state} & \colhead{\footnotesize  (cm$^{-1}$)} & \colhead{\footnotesize  range} & \colhead{\footnotesize    (cm$^{-1}$)} & \colhead{\footnotesize  of labels}  & 
\colhead{\footnotesize  in {\it SEE}} & \colhead{\footnotesize  in {\it HVW}} & \colhead{\footnotesize  in {\it CCS}}  }  
\startdata
CO      &\small  &\small1-2 &\small0-12 &\small2 &\small- &\small667 &\small100 \\
$^{13}$CO      &\small  &\small1-2  &\small0-11 &\small2  &\small2.5 &\small- &\small2.5 \\
C$^{17}$O      &\small  &\small1-2 &\small0-11 &\small2 &\small0.3 &\small- &\small0.3 \\
C$^{18}$O      &\small  &\small1-2 &\small0-11 &\small2 &\small0.2 &\small- &\small0.2 \\
HCO$^+$     &\small   &\small1-3 &\small0-18  &\small3 &\small0.1 &\small0.2 &\small0.1 \\
H$^{13}$CO$^+$     &\small   &\small1-3 &\small0-18  &\small3 &\small- &\small- &\small0.003 \\
HOC$^+$     &\small   &\small    &\small      &\small1 &\small &\small &\small \\
HCN     &\small   &\small1-3 &\small0-18  &\small3  &\small2.0 &\small2.0 &\small1.5 \\
HCN $\nu2$    &\small 712  &\small1-3 &\small0-18  &\small 5  &\small &\small  &\small \\
H$^{13}$CN &\small &\small 1-3 &\small 3-17 &\small3  &\small0.133 &\small0.133 &\small0.033\\
H$^{13}$CN $\nu2$ &\small &\small 1-3 &\small  &\small 5  &\small  &\small  &\small \\
HNC     &\small   &\small1-3 &\small0-18 &\small2 &\small0.2 &\small0.2 &\small0.167 \\
HNC $\nu2$     &\small 477  &\small1-3 &\small3-18 &\small2 &\small &\small &\small \\
HN$^{13}$C   &\small  &\small1-3 &\small0-18 &\small3 &\small0.013 &\small0.01 &\small0.004 \\
H$_2$CO &\small  &\small- &\small0.64 &\small9  &\small- &\small- &\small0.125 \\
SiO &\small &\small2-6 &\small1-30 &\small4 &\small- &\small- &\small0.013 \\
CS &\small &\small2-5 &\small2-35 &\small3  &\small- &\small- &\small0.125 \\
N$_2$H$^+$ &\small &\small1-1  &\small0-3  &\small1 &\small- &\small- &\small0.025 \\
c-C$_3$H$_2$ &\small &\small &\small2-134 &\small67 &\small- &\small- &\small0.36 \\
MgNC &\small &\small7-14 &\small8-36 &\small16  &\small- &\small-  &\small0.10 \\
CH$_3$CN &\small &\small5-15 &\small9-300 &\small 56 &\small0.067 &\small0.100 &\small0.033 \\
CH$_3$CCH &\small &\small5-16  &\small6-315 &\small 65 &\small1.0 &\small2.0 &\small 0.200 \\
CH$_2$CHCN &\small &\small  &\small &\small 120 &\small &\small &\small 0.067 \\
CN &\small &\small1-2 &\small0-11 &\small26 &\small- &\small- &\small4.0 \\
$^{13}$CN &\small &\small1-2 &\small0-11 &\small24 &\small- &\small- &\small0.25 \\
CCH &\small &\small1-3 &\small0-18 &\small17 &\small- &\small- &\small0.5 \\
C$_{3}$N &\small &\small9-22 &\small14-77  &\small26  &\small- &\small- &\small0.118 \\
c-C$_3$H &\small &\small &\small0-41 &\small25  &\small- &\small- &\small0.050 \\
l-C$_3$H &\small &\small &\small5-52  &\small58 &\small- &\small- &\small0.017 \\
C$_4$H &\small &\small9-29 &\small14-142 &\small30 &\small- &\small- &\small 1.0 \\
C$_4$H $\nu_7$ &\small 134 &\small9-22 &\small 12-86 &\small 31 &\small- &\small- &\small  \\
C$_{5}$H &\small &\small- &\small23-100 &\small38 &\small- &\small- &\small0.005 \\
H$_2$C$_{3}$ &\small -&\small -&\small -&  2& - & - & $<$0.005 \\
\hline
\footnotesize TOTAL no-HC$_{2n+1}$N & & & & 666 & & & \\
\hline
H recombination & & & & 44 & & & \\
He recombination & & & & 3 & & & \\
U-lines & & & & 74 & & & \\
\hline
{\bf TOTAL LABELS} & & & & {\bf 3127} & & & \\
\enddata
\label{table-species-bis}
\end{deluxetable}

\begin{deluxetable}{rrrr}
\tablecolumns{4} 
\tablewidth{0pc} 
\tablecaption{Unidentified features in the 80-276 GHz IRAM-30m survey of CRL 618. Frequency given in MHz. Area (A) given in K$\cdot$kms$^{-1}$ and HPLW (W) given in kms$^{-1}$. The $\sigma$ of the feature 
has been established from the following average rms of the data: 4 mK at 3 mm, 8 mK at 2 mm, 11 mK from 204 to 240 GHz, and 14 mk above 240 GHz. Only 74 unidentified 
features arise above the 3$\sigma$ limit.}
\tablehead{ 
\colhead{\small Freq.} & \colhead{\small A (W) $\sigma$} & \colhead{\small Freq.} & \colhead{\small A (W) $\sigma$} \\
\colhead{\small MHz} & \colhead{\small K~km~s$^{-1}$ } & \colhead{\small MHz} & \colhead{\small K~km~s$^{-1}$ } \\
\colhead{ } & \colhead{\small (km~s$^{-1}$) } & \colhead{ } & \colhead{\small (km~s$^{-1}$) } }
\startdata
\small 80559 & \small 0.129 (18) 3$\sigma$    &
\small 80618  &\small 0.069 (7) 3$\sigma$    \\
\small 83659 & \small 0.380 (24) 4$\sigma$                &
\small 85136    &\small0.295 (16) 5$\sigma$ \\ 
\small 87013  & \small0.160 (17) 3$\sigma$ &   
\small 89120    &\small 0.124 (9) 4$\sigma$    \\ 
\small 89600$^1$ & \small 0.239 (27) 3$\sigma$ &  
\small 89820 & \small 0.124 (9) 4$\sigma$   \\
\small 91156  & \small -0.320 (20) 4$\sigma$  &  
\small 91238$^2$  & \small-0.082 (5) 4$\sigma$ \\ 
\small 91383  & \small 0.389 (21) 5$\sigma$  &  
\small 91441$^2$  & \small 0.100 (9)  3$\sigma$ \\ 
\small 92414$^2$ & \small -0.105 (4) 7$\sigma$       & 
\small 92814$^3$ & \small -0.147 (7) 5$\sigma$ \\ 
 \small 95667  & \small 0.289 (23) 3$\sigma$ &  
\small 97637  & \small0.447 (48) 3$\sigma$ \\
\small 97728    &\small0.167 (17) 3$\sigma$    &
\small 97772  &\small 0.246 (22) 3$\sigma$ \\
\small 98822  & \small0.301 (27) 3$\sigma$   &
\small 98856  & \small0.392 (34) 3$\sigma$  \\
\small 100265  &\small 0.304 (18) 4$\sigma$     & 
\small 100540 & \small0.530 (37) 4$\sigma$    \\
\small 101372 & \small -0.115 (5) 6$\sigma$    &
\small 101751   &\small0.212 (14) 4$\sigma$     \\ 
\small 102222   &\small 0.261 (18) 4$\sigma$   &
\small 103426   &\small 0.160 (17) 3$\sigma$    \\
\small 103910$^4$   &\small0.237 (19) 3$\sigma$   & 
\small 107121$^5$   &\small 0.281 (24) 3$\sigma$ \\ 
\small 107275   &\small0.310 (25) 3$\sigma$ &   
\small 107620   &     \small 0.377 (31)   3$\sigma$            \\
\small 115426   &\small 0.675 (19) 9$\sigma$   &        
\small 132378   &\footnotesize 0.22/-0.49 (9/27) 4$\sigma$ \\   
\small  132679 & \small 0.735 (52) 3$\sigma$   &
\small  134239  &\small0.248 (11) 3$\sigma$      \\ 
\small  134525  &  \small 0.488 (28)  3$\sigma$      &
\small  137291  &\small -0.121 (4) 4$\sigma$ \\   
\small 137720   &\small 0.446 (15) 4$\sigma$    &
\small  141811 &\small 0.117 (8) 3$\sigma$  \\
\small  142293  &\small 0.165 (9) 3$\sigma$   & 
\small 146210   &\small 0.445 (17) 3$\sigma$    \\ 
\small 150349 & \footnotesize -0.054/-0.055 (3/6) 3$\sigma$ & 
\small 150620 & 0.470 (42) 3$\sigma$ \\ 
\small 160670 & 0.835 (27) 4$\sigma$ &   
\small  160952$^6$   &\small 3.669 (56) 9$\sigma$   \\  
\small 170642   &\small 0.847 (35) 3$\sigma$    &
\small 172998   &\small-0.092 (4)  3$\sigma$    \\ 
\small  173106  &\small-0.389 (15) 3$\sigma$   & 
\small 173572   &\small -0.130 (6) 3$\sigma$ \\   
\small 175840   &\small 0.057 (2) 3$\sigma$   &
\small 205003$^7$   &  -  \\  
\small 218042   &\small 1.144 (17) 6$\sigma$  &    
\small  218207  &\small1.640 (17) 9$\sigma$  \\
\small 219270   &\small 0.575 (18) 3$\sigma$ &   
\small 220431   &\small 0.556 (17) 3$\sigma$   \\
\small 221366   &\small 2.132 (59) 3$\sigma$  &    
\small 221980$^7$   &\small 3.612 (28) 11$\sigma$ \\ 
\small 228507   &\small 0.105 (3)  3$\sigma$    & 
\small 234936   &\small 0.693 (11) 5$\sigma$ \\
\small 235597   &\small 0.422 (16) 3$\sigma$ &   
\small  235606  &\small0.152 (6) 3$\sigma$   \\
\small 241337$^8$   &\small 3.799 (45) 7$\sigma$ & 
\small  241388$^8$  &\small 1.175 (19) 5$\sigma$ \\ 
\small 241498   &\small 0.913 (15) 5$\sigma$  &  
\small  242642  &\small 0.178 (5) 3$\sigma$  \\
\small 245120   &\small 0.690 (17) 3$\sigma$   & 
\small 248647   &\small 0.139 (4)  3$\sigma$  \\
\small  249111  &\small0.487 (17) 3$\sigma$ &  
\small 256428   &\small 0.238 (4) 5$\sigma$   \\
\small 258314   &\small 0.811 (15) 4$\sigma$  &    
\small 260295   &\small 0.293 (6) 4$\sigma$ \\
\small 260426   &\small 0.233 (5)  3$\sigma$   &  
\small  263475  &\small0.328 (8) 3$\sigma$    \\
\small 263840   &\small 0.186 (4) 3$\sigma$  &  
\small  272988  &\small1.560 (11) 10$\sigma$ \\
\hline
\multicolumn{2}{l}{$^1$ \footnotesize (id) HC$^{13}$CCCCN J=34-33} &
\multicolumn{2}{l}{$^2$ \footnotesize (cm) only one channel} \\
\multicolumn{2}{l}{$^3$ \footnotesize (id) C$_3$H$_2$} &
\multicolumn{2}{l}{$^4$ \footnotesize (id) H 56$\gamma$} \\
\multicolumn{2}{l}{$^5$ \footnotesize (id) HC$_5$N 3$\nu_{10}$(1+) J=40-39} &
\multicolumn{2}{l}{$^6$ \footnotesize (id) CH$_2$CN J=8-7} \\
\multicolumn{2}{l}{$^7$ \footnotesize (bl) CH$_3$CCH} &
\multicolumn{2}{l}{$^8$ \footnotesize (id) CH$_2$CN J=12-11} \\
\multicolumn{4}{l}{\footnotesize id: possible identification; cm: comment; bl: existing blending} \\
\enddata
\label{u-lines}
\end{deluxetable}


\end{document}